\documentclass[journal,twoside,web]{ieeecolor}
\usepackage{generic}
\usepackage{cite}
\usepackage{amsmath,amssymb,amsfonts}
\usepackage{algorithmic}
\usepackage{graphicx}
\usepackage{textcomp}
\usepackage{tablefootnote}
\def\BibTeX{{\rm B\kern-.05em{\sc i\kern-.025em b}\kern-.08em
    T\kern-.1667em\lower.7ex\hbox{E}\kern-.125emX}}
\usepackage{etoolbox}
\makeatletter
\@ifundefined{color@begingroup}%
  {\let\color@begingroup\relax
   \let\color@endgroup\relax}{}%
\def\fix@ieeecolor@hbox#1{%
  \hbox{\color@begingroup#1\color@endgroup}}
\patchcmd\@makecaption{\hbox}{\fix@ieeecolor@hbox}{}{\FAILED}
\patchcmd\@makecaption{\hbox}{\fix@ieeecolor@hbox}{}{\FAILED}

\newcommand{\Changes}[1]{\textcolor{black}{#1}}

\usepackage{hyperref}

\begin{document}

\title{Latent Similarity Identifies Important Functional Connections for Phenotype Prediction}
\author{Anton Orlichenko, Gang Qu, Gemeng Zhang, Binish Patel, Tony W. Wilson, Julia M. Stephen, Vince D. Calhoun, \IEEEmembership{Fellow, IEEE,} and Yu-Ping Wang, \IEEEmembership{Senior Member, IEEE} \\
\thanks{Manuscript received on September 14, 2022. This work was supported in part by NIH grants (P20GM109068, P20GM144641, R01MH121101, R01MH104680, R01MH107354, R01MH103220, R01EB020407, R56MH124925) and NSF grant (\#1539067). \textit{(Corresponding author: Yu-Ping Wang.)}}
\thanks{Anton Orlichenko, Gang Qu, Gemeng Zhang, Binish Patel, and Yu-Ping Wang are with the  Department of Biomedical Engineering, Tulane University, New Orleans, LA 70118. (e-mail: wyp@tulane.edu).}
\thanks{Julia M. Stephen is with the Mind Research Network, Albuquerque, NM 87106. (e-mail: jstephen@mrn.org).}
\thanks{Tony W. Wilson is with the Institute for Human Neuroscience, Boys Town National Research Hospital, Boys Town, NE 68010. (e-mail: tony.wilson@boystown.org).}
\thanks{Vince D. Calhoun is with the Tri-Institutional Center for Translational Research in Neuroimaging and Data Science (TReNDS) (Georgia State University, Georgia Institute of Technology, Emory University), Atlanta, GA 30303. (e-mail: vcalhoun@gsu.edu).}
\thanks{Copyright (c) 2021 IEEE. Personal use of this material is permitted. However, permission to use this material for any other purposes must be obtained from the IEEE by sending an email to pubs-permissions@ieee.org.}}

\maketitle

\urlstyle{same}

\begin{abstract}
\textit{Objective}: Endophenotypes such as brain age and fluid intelligence are important biomarkers of disease status. However, brain imaging studies to identify these biomarkers often encounter limited numbers of subjects but high dimensional imaging features, hindering reproducibility. Therefore, we develop an interpretable, multivariate classification/regression algorithm, called Latent Similarity (LatSim), suitable for small sample size but high feature dimension datasets. \textit{Methods}: LatSim combines metric learning with a kernel similarity function and softmax aggregation to identify task-related similarities between subjects. Inter-subject similarity is utilized to improve performance on three prediction tasks using multi-paradigm fMRI data. A greedy selection algorithm, made possible by LatSim's computational efficiency, is developed as an interpretability method. \textit{Results}: LatSim achieved significantly higher predictive accuracy at small sample sizes on the Philadelphia Neurodevelopmental Cohort (PNC) dataset. Connections identified by LatSim gave superior discriminative power compared to those identified by other methods. We identified 4 functional brain networks enriched in connections for predicting brain age, sex, and intelligence. \textit{Conclusion}: We find that most information for a predictive task comes from only a few (1-5) connections. Additionally, we find that the default mode network is over-represented in the top connections of all predictive tasks. \textit{Significance}: We propose a novel prediction algorithm for small sample, high feature dimension datasets and use it to identify connections in task fMRI data. Our work can lead to new insights in both algorithm design and neuroscience research.
\end{abstract}


\section{Introduction}
\IEEEPARstart{F}{unctional} magnetic resonance imaging (fMRI) provides a non-invasive estimate of brain activity by exploiting the blood oxygen level-dependent (BOLD) signal \cite{Belliveau1991FunctionalMO}. This high-acuity imaging data can be used to predict variables like age, sex, intelligence, and disease status \cite{10.1117/12.2613172}\cite{ICER2020105444}\cite{9428628}\cite{Du2018ClassificationAP}. Interestingly, the gap between fMRI-predicted brain age and biological age can identify Alzheimer's disease patients prior to the onset of symptoms \cite{Millar2022-ov}. Prediction is hindered, however, by the combination of small sample size and very high feature number. This results in models that have poor reproducibility and generalizeability \cite{Berisha2021-bz}.

Studies with small sample size only have the power to detect very large effects. Many effects that are found in small studies may be due to noise. When identifying regions that are associated with in-scanner tasks, it was found that the average minimum cohort size needed to reproducibly identify the same region 50\% of the time in independent samples was N=36 \cite{Turner2018-vq}. In contrast, models deployed clinically use thousands of subjects for training and validation \cite{Salehinejad2021ARD}. In 2017 and 2018, the median cohort sizes for published experimental and clinical MRI studies were 23 and 24 subjects, respectively, and less than 1\% of the 272 papers surveyed reported cohort sizes greater than 100 \cite{SZUCS2020117164}. This may be attributed to both cost, at \$500-\$1000 per subject, and the difficulty of collecting the data, stemming from long scan times, subject discomfort in the scanner, and experimental design \cite{SZUCS2020117164}.

Additionally, for fMRI-based predictions to be useful clinically, they must be interpretable. There is a large literature on the interpretability of machine learning in medical imaging \cite{LUNDERVOLD2019102}\cite{SALAHUDDIN2022105111}; however, there is often a tradeoff between model accuracy and interpretability. This raises questions about robustness in the clinical setting \cite{Varoquaux2022-et}. For example, \textit{Zhang et al.} show that different processing methods can yield similar accuracy in a sex prediction task, but with different discriminative features identified by each method \cite{Zhang2020GenderDA}. Identifying a minimal set of valid functional connections may increase model robustness, and make inroads into causal analysis of brain networks \cite{pmid30793082}. 

Finally, many recent studies in the deep learning field shift their focus to integrate data from multiple omics \cite{Subramanian2020-ep}, or multiple omics and imaging \cite{Hu2021InterpretableMF}. This is done for two purposes: to improve prediction accuracy and to learn novel interactions between different modalities. CCA-based models have been proposed that use response variable-guided feature alignment \cite{Gross2015-kt} \cite{9767566}. However, these models do not consider inter-subject relationships and cannot control disentanglement between different predictive tasks.

\begin{figure}
	\centering
	\includegraphics[width=8cm]{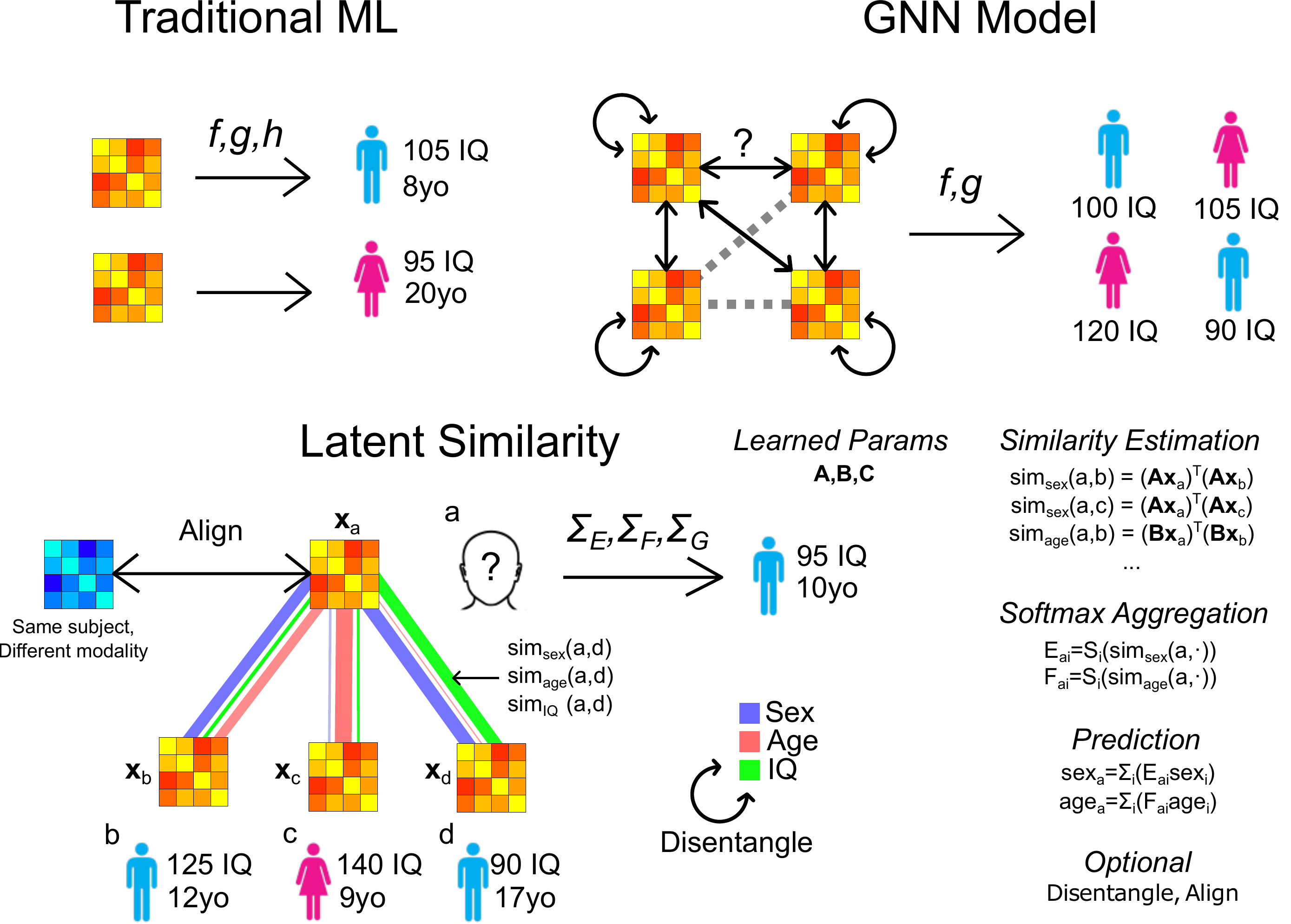}
	\caption{\Changes{An overview of the Latent Similarity model. In traditional ML, estimation of response variables is decoupled from inter-subject similarity, whereas GNN models require additional degrees of freedom to estimate edges between subjects. Our model calculates similarity between subjects based on a set of response variables and incorporates multi-modal feature alignment (in addition to ensembling) as well as sparsity and feature disentanglement.}}
	\label{fig:overview}
\end{figure}

In this paper, we introduce LatSim (Figure~\ref{fig:overview}), a model in the spirit of metric learning \cite{Kaya2019DeepML}, that is both robust and interpretable. Traditional machine learning (ML) models in fMRI, which work directly on functional connectivity (FC) \cite{van2010exploring}, are vulnerable to noise or random confounders like scanner drift or head motion \cite{Kato2020-hu}. Graph neural networks (GNN) use inter-subject information as an adjunct to calculations performed directly on FC \cite{Zhou2020GraphNN}. However, graph edges may be ambiguous or non-binary, requiring additional degrees of freedom for their estimation \cite{Velickovic2018GraphAN}\cite{NEURIPS2018_53f0d7c5}. In contrast, LatSim learns an inter-subject similarity metric, $d(\mathbf{x}_a, \mathbf{x}_b)$, and uses the inter-subject similarity, without a self-loop, to make predictions.


The contribution of our work is three-fold. First, we propose a novel metric learning-based model, LatSim, which is robust, interpretable, computationally efficient, multi-view, and multi-task. Second, we use LatSim and a greedy selection algorithm to identify the most discriminitive connections for age, sex, and intelligence prediction among adolescents in the Philadelphia Neurodevelopmental Cohort (PNC) dataset  \cite{Satterthwaite2014NeuroimagingOT}. We show that the such connections are superior to those identified by saliency maps. Third, we give a justification why LatSim performs better than traditional ML models with low sample sizes and high feature dimensionality.

The rest of this paper is organized as follows. Section~\ref{sec:methods} gives the mathematical foundations of LatSim and its relationship to other models. Section~\ref{sec:results} provides simulation and experimental results. Section~\ref{sec:discussion} discusses significant brain networks and reasons why LatSim performs better in the low sample-size, high-dimensionality regime. Section~\ref{sec:conclusion} concludes with a recapitulation of the work. We make the code publicly available at the link in the footnote.\footnote{\url{https://github.com/aorliche/LatentSimilarity/}.}

\begin{table}
    \centering
    \caption{Commonly used notation.}
    \begin{tabular}{|c|l|}
        \hline
        Notation & \multicolumn{1}{|c|}{Description} \\
        \hline
        $\mathbf{X} \in \mathbb{R}^{N\times d}$ & A matrix of dimension $N$ by $d$\\
        $X_{ij}$ & The ($i$,$j$)-th entry of matrix $\mathbf{X}$ \\
        $\mathbf{X}_{i,:}$ & The $i$-th row of matrix $\mathbf{X}$ \\
        $\mathbf{X}_i,\mathbf{X}^{(i)}$ & The $i$-th matrix in a set of matrices \\
        $\mathbf{X}^\text{T}$ & The transpose of matrix $\mathbf{X}$ \\
        $A,B$ & Random variables \\
        $F_i$ & The $i$-th element of a set \\
        $y_i$ & The $i$-th entry of vector $y$\\
        $\odot$ & The Hadamard product \\
        $\mathbf{1}$ & A matrix of ones \\
        $\text{diag}(\mathbf{a})$ & A square matrix with the elements of \\
        & $\mathbf{a}$ on the main diagonal, 0s elsewhere \\
        $\Sigma_{abc}$ & Summation over indices $a,b,c$ \\
        $\mathbb{E}[\cdot]$ & Expectation \\
        $\text{Var}[\cdot],\text{Cov}[\cdot]$ & Variance, covariance \\
        $\mu, \sigma^2, \rho$ & Mean, variance, correlation \\
        $|C|$ & Cardinality of set $C$ \\
        $||\cdot||_1$ & The $l_1$ norm \\
        $||\cdot||_2$ & The $l_2$ norm \\
        \hline
    \end{tabular}
    \label{tab:definitions}
\end{table}

\section{Methods}
\label{sec:methods}

\subsection{Kernel CCA}
\label{subsec:kcca}

To compute similarity between subjects, we utilize ideas from canonical correlation analysis (CCA) \cite{10.1093/biomet/28.3-4.321}\cite{LI2020105073}. Conventional CCA seeks to find relationships between the features of two different views of a dataset. It aligns the two views, $\mathbf{X}_1 \in \mathbb{R}^{N\times d_1}$ and $\mathbf{X}_2 \in \mathbb{R}^{N\times d_2}$, by finding canonical variables $\mathbf{w}_1$ and $\mathbf{w}_2$ that maximize the correlation between $\mathbf{X}_1\mathbf{w}_1$ and $\mathbf{X}_2\mathbf{w}_2$:

\begin{equation}
    \begin{split}
        \underset{\mathbf{w}}{\text{maximize}} \quad & \mathbf{w}^\text{T}_1\mathbf{X}^\text{T}_1\mathbf{X}_2\mathbf{w}_2 \\
        \text{s.t.} \quad & \mathbf{w}^\text{T}_1\mathbf{X}^\text{T}_1\mathbf{X}_1\mathbf{w}_1 = 1, \\
        & \mathbf{w}^\text{T}_2\mathbf{X}^\text{T}_2\mathbf{X}_2\mathbf{w}_2 = 1
    \end{split}
\end{equation}

\noindent where $N$ is the number of subjects and $d_1=d_2=d$ is the feature dimension. Kernel CCA (kCCA) \cite{DBLP:journals/corr/abs-cs-0609071} \cite{7822570} transforms features into a reproducing kernel Hilbert space (RKHS), and finds the alignment between the transformed features $\mathbf{K}_1$ and $\mathbf{K}_2$. The similarity in the RKHS is $k(\mathbf{X}_{i,:},\mathbf{X}_{j,:}) = \phi(\mathbf{X}_{i,:})^T\phi(\mathbf{X}_{j,:})$, where $\phi: \mathbb{R}^d \mapsto \mathbb{R}^{d'}$ is the feature transformation. LatSim learns a linear kernel $\mathbf{A} \in \mathbb{R}^{d\times d'}$; however, this still allows detection of nonlinear relationships.

\Changes{The main idea behind CCA and kCCA is to maximize the similarity between two or more signals after some constrained transformation. This constrained transformation moves the data to a latent space, which may be of lower dimension. The limitation of CCA and kCCA is that they are unsupervised learning techniques that must account for every similarity between the signals, not just those relevant for a particular application, although recent work is tackling this problem \cite{9767566}.}

\subsection{Latent similarity}
\label{subsec:latsim}

\Changes{In contrast to unsupervised learning, LatSim maximizes similarity of subjects relative to a response variable of interest, such as age, sex or intelligence.} First, similarities are computed as the inner product of the low-dimensional projections of subject features, based on a learned kernel: 

\begin{equation}
    \begin{split}
    \Changes{\text{sim}(a,b)} &= \Changes{\langle \phi(\mathbf{x}_a), \phi(\mathbf{x}_b) \rangle }\\
     &=\Changes{ \mathbf{x}_a\mathbf{A}\mathbf{A}^\text{T}\mathbf{x}^\text{T}_b,}
    \end{split}
\end{equation}

\noindent where $\mathbf{A} \in \mathbb{R}^{d\times d'}$ is the kernel matrix and $\mathbf{x}_a, \mathbf{x}_b \in \mathbb{R}^{d}$ are feature vectors for subjects $a$ and $b$, respectively. These similarities are then adjusted by passing them through a softmax activation function while masking each subject's self-similarity. The entire model for a single predictive task and a single fMRI paradigm is as follows:

\begin{equation}
    \begin{split}
        \mathbf{M}&=\text{diag}(\mathbf{\infty}),\\
        \mathbf{E}&=S_{Row}((\mathbf{1}-\mathbf{M})\odot \mathbf{X}\mathbf{A}\mathbf{A}^\text{T}\mathbf{X}^\text{T}),\\
        S(\mathbf{z})_i&=\frac{e^{z_i/\tau}}{{\Sigma^N_{j=0}}e^{z_j/\tau}}, \\
    \end{split}
\end{equation}

\noindent where $\mathbf{E} \in \mathbb{R}^{N\times N}$ is the final similarity matrix, $\mathbf{M} \in \mathbb{R}^{N\times N}$ is a mask to remove self-loops in predictions, $\mathbf{\infty} \in \mathbb{R}^N$ is a vector of infinite-valued elements, $\mathbf{1} \in \mathbb{R}^{N\times N}$ is a matrix of ones, $\mathbf{X} \in \mathbb{R}^{N\times d}$ is the feature matrix, $\mathbf{A} \in \mathbb{R}^{d\times d'}$ is the kernel taking connectivity features to a lower latent dimension, $N$ is the number of subjects, $d$ is the number of features (FCs), $S(\mathbf{z})_i$ is the softmax function with temperature $\tau$, and $S_{Row}(\mathbf{Z})$ is a function applying softmax to each row of the input matrix. High or low temperature $\tau$ determines whether the subject-subject similarity matrix $\mathbf{E}$ is more dense or sparse, respectively. The final similarity matrix of training and test set subjects is multiplied by the training set response variable to yield the prediction:

\begin{equation}
    \mathbf{\hat{y}} = \mathbf{E}\mathbf{y}_{train}
\end{equation}

The model is trained, using gradient descent, by minimizing the following objective function. Here we assume for brevity the existence of two fMRI feature matrices $\mathbf{X}_a$ and $\mathbf{X}_b$, and two predictive tasks, one regression (1) and one classification (2), for which we identify four kernel matrices $\mathbf{A}_{1a}$, $\mathbf{A}_{1b}$, $\mathbf{A}_{2a}$ and $\mathbf{A}_{2b}$:

\begin{equation}
\label{eq:costfn}
    \begin{split}
        & \underset{\mathbf{A}_{1a},\mathbf{A}_{1b},\mathbf{A}_{2a},\mathbf{A}_{2b}}{\text{minimize}} \\
        & \qquad \frac{1}{N}(\mathbf{y}^{(1)}-\mathbf{E}^{(1a)}\mathbf{y}^{(1)})^2 + \\
        & \qquad \frac{1}{N}(\mathbf{y}^{(1)}-\mathbf{E}^{(1b)}\mathbf{y}^{(1)})^2 + \\
        & \qquad \gamma_1\frac{1}{N}\Sigma^N_{n=1}\Sigma^C_{c=1}\mathbf{Y}^{(2)}_{:,c}\cdot \text{log}(\mathbf{E}^{(2a)}\mathbf{Y}^{(2)})_{:,c} + \\
        & \qquad \gamma_2\frac{1}{N}\Sigma^N_{n=1}\Sigma^C_{c=1}\mathbf{Y}^{(2)}_{:,c}\cdot \text{log}(\mathbf{E}^{(2b)}\mathbf{Y}^{(2)})_{:,c} + \\
        & \qquad \lambda_1||\mathbf{A}_{1a}||_1 + \lambda_2||\mathbf{A}_{1b}||_1 + \\
        & \qquad \lambda_3||\mathbf{A}_{2a}||_1 +
        \lambda_4||\mathbf{A}_{2b}||_1 +\\
        & \qquad \alpha_1||\mathbf{A}_{1a}\odot\mathbf{A}_{2a}||_1 + \\
        & \qquad \alpha_2||\mathbf{A}_{1b}\odot\mathbf{A}_{2b}||_1
        + \\
        & \qquad \beta_1||\mathbf{X}_a\mathbf{A}_{1a}-\mathbf{X}_b\mathbf{A}_{1b}||_2 + \\
        & \qquad \beta_2||\mathbf{X}_a\mathbf{A}_{2a}-\mathbf{X}_b\mathbf{A}_{2b}||_2, \\
    \end{split}
\end{equation}

\noindent where $\mathbf{E}^{(1a)} \in \mathbb{R}^{N\times N}$, for example, is the similarity matrix for task 1 and fMRI paradigm $a$, $\mathbf{y}^{(1)} \in \mathbb{R}^N$ (numeric) and $\mathbf{Y}^{(2)} \in \mathbb{R}^{N\times C}$ (one-hot categorical) are the stacked response variables for tasks 1 and 2, respectively, $N$ is the number of subjects, $C$ is the number of classes in task 2, $\gamma_i$ is a task importance weight, $\lambda_i$ is a sparsity-inducing hyperparameter, $\alpha_i$ is a hyperparameter promoting feature disentanglement, and $\beta_i$ is a hyperparameter promoting alignment between fMRI paradigms. \Changes{Note that our experiments on the PNC dataset in Section~\ref{subsubsec:dataset} used precomputed vectorized functional connectivity matrices as the input, e.g., $\mathbf{X}$ is a matrix where each row is the vectorized FC of one subject.}

In the conventional image domain, \textit{Zheng et al.} have proposed a similar metric learning approach using softmax aggregation for image classification \cite{Zheng_2022_CVPR}. However, their work makes use of a pre-trained backbone, is semi-supervised, and does not provide all of the possibilities for feature selection, disentanglement, and alignment as does LatSim (see Equation~\ref{eq:costfn}).

\begin{figure}
    \centering
    \includegraphics[width=8cm]{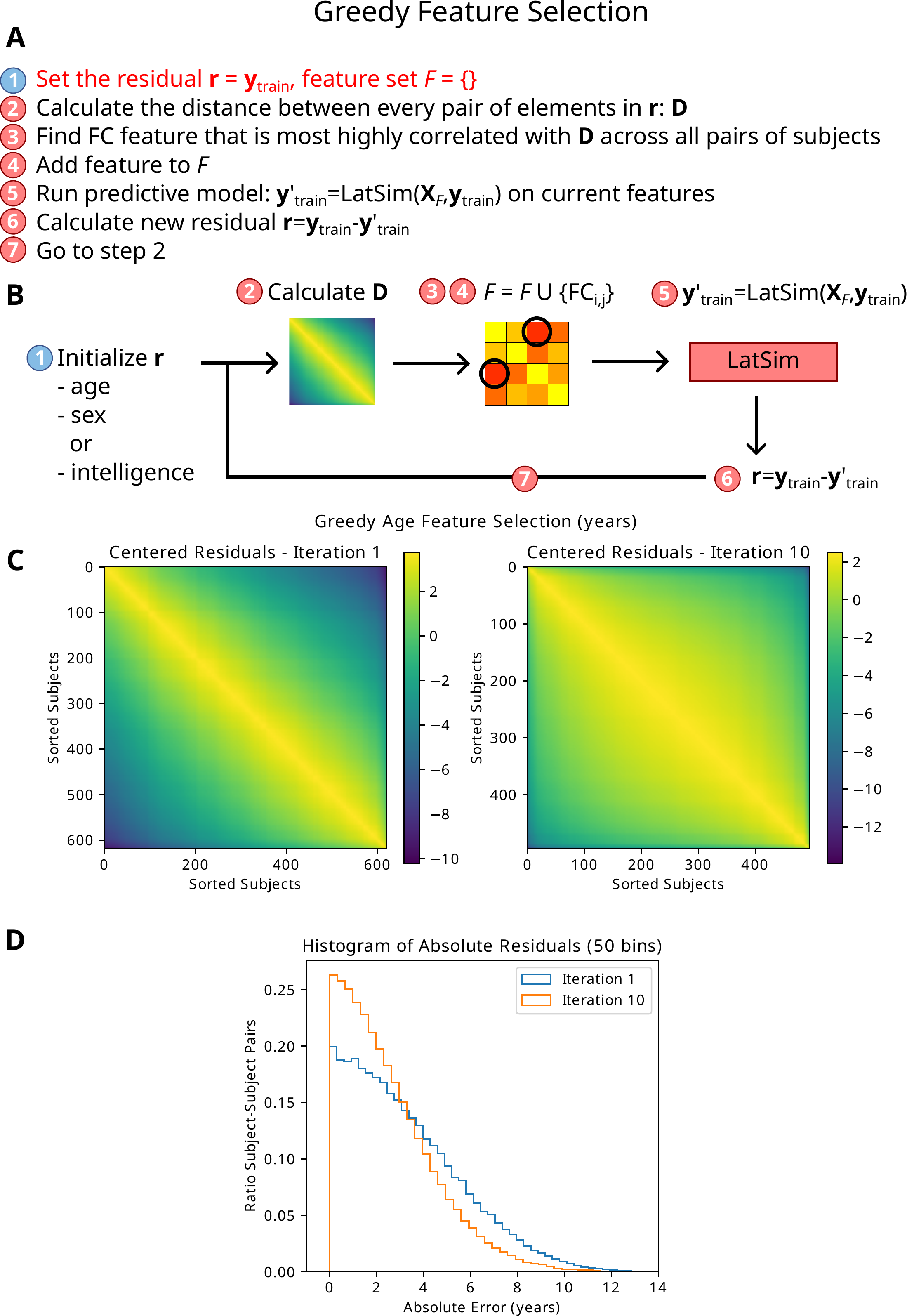}
    \caption{\Changes{The greedy feature selection algorithm. \textbf{A.} A summary of the algorithm. \textbf{B.} A flowchart representation. \textbf{C.} Visualization of the residual distance matrices used to choose an FC feature at each iteration, at iterations 1 and 10. \textbf{D.} Histogram of absolute residual distance matrix values at iterations 1 and 10. Since LatSim works on subject pairs, our objective is to fit distances between residuals.}}
    \label{fig:greedyalgo}
\end{figure}

\subsection{Greedy selection algorithm and model interpretability}
\label{subsec:interp}

A greedy selection algorithm was developed to compare with other interpretability methods \cite{pmlr-v44-atzmon2015}. The algorithm selects connections one at a time by ranking their ability to separate dissimilar subjects, i.e., their ability to minimize similarity between subjects that are ``far apart" with regards to the current residual:

\begin{equation}
    \label{eq:greed}
    \begin{split}
        \mathbf{r}^{(i)} &= \text{LatSim}(\mathbf{X}_{F_{i-1}}, \mathbf{y}) - \mathbf{y},\\
        D_{ab} &= (r^{(i)}_a-r^{(i)}_b)^2,\\
        \mathbf{D} &= \mathbf{D} - \frac{1}{N^2}\Sigma_{ab}D_{ab},\\
        F_i &= F_{i-1}\cup\{\underset{j}{\text{argmin}}\;\Sigma_{ab}\;  (D_{ab}X_{aj}X_{bj})\},\\
    \end{split}
\end{equation}

\noindent where $\text{LatSim}: \mathbb{R}^{N\times |F_{i-1}|} \to \mathbb{R}^N$ is the predictive model, $r_a^{(i)}$ is the residual at iteration $i$ for subject $a$, $\mathbf{D} \in \mathbb{R}^{N \times N}$ is a centered matrix of differences between residuals, $F_i = \{0,\ldots,i\}$ is the set of selected connections at iteration $i$, $\mathbf{X} \in \mathbb{R}^{N \times d}$ is the vectorized FC matrix for all subjects, and $\mathbf{y} \in \mathbb{R}^{N}$ is the response variable. \Changes{A summary of the algorithm is presented in Figure~\ref{fig:greedyalgo}. We describe feature selection results in Section~\ref{subsubsec:sigfcs}.}

The greedy algorithm can select the several dozen most relevant features given a single predictive task. To select discriminative features using the fully trained model, we find the correlation between subject similarities and residual distances, as in Equation~\ref{eq:greed} above, except the FCs are multiplied by the learned model weights:

\begin{equation}
    \label{eq:select}
    F = \underset{j}{\text{argsort}} \; \Sigma_{abd}\; (D_{ab} A^2_{dj}X_{aj}X_{bj}),\\
\end{equation}

\noindent where the residual is set to the response variable, $\mathbf{D}$ is calculated as before, $\mathbf{A} \in \mathbb{R}^{d\times d'}$ is the set of model weights, and $F$ is the resulting set of ranked features. 

\Changes{Except for greedy feature selection, we optimized prediction of all three response variables (age, sex, and intelligence) at the same time in the same LatSim model. Greedy selection required optimizing a single task at once, as the best feature for age prediction may not be the best feature for sex or intelligence prediction.} LatSim was trained using PyTorch on an NVIDIA Titan Xp with CUDA support.

\subsection{Spurious correlation}

\Changes{We hypothesize that overfitting occurs due to feature noise or confounds, such as scanner motion, whose effects are more severe for smaller size cohorts. These confounds may create spurious correlations in a subset of the cohort.}

\Changes{We define a spuriously correlated feature $X$ to be one that appears to be highly correlated with response variable $Y$ for only a subset of subjects:}

\begin{equation}
    |\rho_{S}| \begin{cases}
    > 0 & s\in S \\
    \approx 0 & s\in C\setminus S
    \end{cases}
\end{equation}

\noindent \Changes{where $\rho_S$ is the value of the spurious correlation, $C$ is the study cohort, and $S\subseteq C$ is a subset of the cohort such that $C\setminus S$ is maximized.}

\Changes{
Note that spurious correlation may actually be true correlation identifying subgroups, but we hypothesize that a spurious correlation is more likely to be false as $|S|$ decreases. We conduct simulation experiments in Section~\ref{subsec:simulation} that suggest LatSim is more robust against spurious correlation than traditional feature-based models. When $|S|$ is close to $|C|$, and the effect is systematic, we cannot tell whether the correlation is true or false.}

\section{Results}
\label{sec:results}

\begin{figure}
    \centering
    \includegraphics[width=8cm]{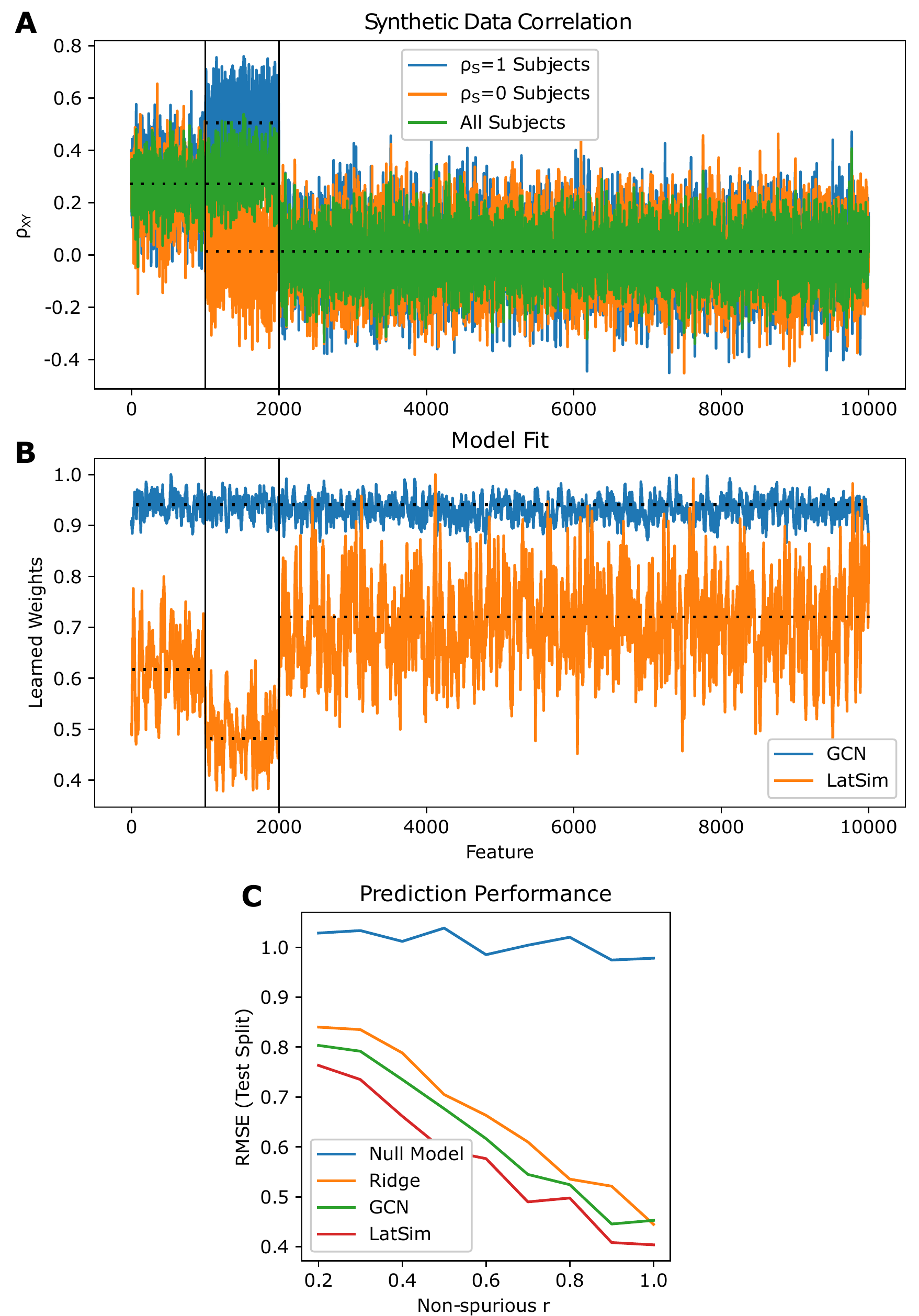}
    \caption{Results of simulations on synthetic data with spurious correlation. \Changes{\textbf{A.} Data generated with non-spurious $\rho=1/(2\sqrt{4.25})\approx0.25$, present in all subjects, and spurious $\rho_S\approx1/\sqrt{5}\approx0.5$, present in half of subjects.} Correlation of response variable with feature for the training set is shown. Only the first two thousand features have any information relevant for prediction. \Changes{\textbf{B.} Absolute value of learned model weights for the GCN and LatSim models, averaged over the first hidden layer (GCN) or latent dimension (LatSim). Weights are smoothed by a convolution kernel of size 20 to aid visualization. \textbf{C.} Average predictive performance (RMSE between ground truth $y_i$ and predicted $\hat{y}_i$) over 6 independent train/validation/test splits, evaluated on the test split.}}
    \label{fig:simulation}
\end{figure}

We first demonstrate the superior performance of LatSim in a simulation study, then apply it to brain development fMRI data consisting of children and adolescents. We use both full-model and greedy feature selection to identify important functional connections for age, sex, and intelligence prediction.

\subsection{Simulation experiment}
\label{subsec:simulation}

We performed a simulation experiment to test LatSim in the presence of a ground truth dataset. A set of $N_{train}=40$, $N_{val}=120$, and $N_{test}=120$ subjects with 10,000 normally-distributed features \Changes{$x_{ni}$ was generated, where $n$ and $i$ refer to subject and feature, respectively. Each subject was also associated with a response variable $y_n$. The data generation process for each subject was as follows:}

\begin{equation}
    \begin{split}
        & y_n\;\sim\mathcal{N}(0,1), 
         \qquad z_{ni}\sim\mathcal{N}(0,4) \\
        x_{ni} &= \begin{cases}
            z_{ni}+y_nr, & \text{if $i\leq1000$}\\
            z_{ni}+y_nr_S, & \text{if $1000<i\leq2000$ and $n$ even}\\
            z_{ni}, & \text{otherwise}
          \end{cases}
    \end{split}
\end{equation}

\noindent \Changes{where $r$ and $r_S$ are correlation-generating parameters for non-spurious and spurious correlations, respectively. In other words, the first 1,000 features were correlated with the response variable at level $\rho$, the next 1,000 features of half of the subjects were correlated at level $\rho_S$ (and had 0 correlation for the other half of subjects), and the remaining 8,000 features were left uncorrelated. We varied $r$ from 0.2 to 1 while keeping $r_S=1$. It can be seen that final feature to response variable correlation is $\rho=r/\sqrt{r^2+4}$ for correlated features for all subjects, and $\rho_S\approx r_S/\sqrt{r^2_S+4}$ for spuriously correlated features for half of subjects.}

The simulation showed that LatSim performs better than both a GCN \cite{Kipf2017SemiSupervisedCW} and Ridge Regression model in the presence of the spurious correlation $\rho_S$ (see Figure~\ref{fig:simulation}). \Changes{Additionally, LatSim was the only model identifying the three types of features: correlated, spuriously correlated, and uncorrelated. All results are on the test split.} We believe insensitivity to spurious correlation is one of the reasons that LatSim performs well in the low-sample, high-dimensionality regime (see Section~\ref{subsec:robustness}). A multi-layer perceptron (MLP) with L1-regularization performed as well as Ridge Regression (not shown). The GCN model was not interpretable via either weight magnitude or gradient-based saliency. The MLP model identified only sparse features and selected features in the non-informative range. In contrast, LatSim was able to consistently identify the full range of informative features.

Notably, the weights are smaller for correlated features than for non-correlated features. This is an artifact of taking the absolute value of weights in order to average them across latent dimensions. Conversely, the spuriously-correlated weights are, on average, smaller than the constantly-correlated weights. To explain, suppose there are 2 sets of features, $A$ and $B$, which are correlated and non-correlated, respectively. The similarity between two subjects will be:

\begin{equation}
    \begin{split}
        \mathbb{E}[(w_AA_1&+w_BB_1)(w_AA_2+w_BB_2)] \\  &=w^2_A\mathbb{E}[A_1A_2] + w^2_B\mathbb{E}[B_1B_2] \\
        &\quad\quad + w_Aw_B\mathbb{E}[A_1B_2] + w_Aw_B\mathbb{E}[A_2B_1] \\
        &= w^2_A\mathbb{E}[A_1A_2] > 0,
    \end{split}
\end{equation}

\noindent hence it doesn't matter what magnitude the weights $B$ have, because the expectation of the non-$A_1A_2$ terms is zero due to independence and the standard normal distribution of features. Conversely, if there is a subset of features $A$ that are spuriously-correlated, it is beneficial to reduce the spurious weights compared to the non-spurious ones.

\subsection{Brain development study}

\subsubsection{Dataset}
\label{subsubsec:dataset}

\begin{table}
	\caption{Demographic information for the subset of the PNC dataset used in our experiments. \Changes{WRAT score has been adjusted from its raw value by regressing out the effects of age.}}
	\centering
	\begin{tabular}{|c|c|}
	    \hline
		& Number of Subjects \\
	    \hline
		Males & 286 \\
		Females & 334 \\
		Total & 620 \\
		\hline
	\end{tabular}\\
	\vspace{0.3cm}
	\begin{tabular}{|l|c|c|c|}
	\hline
	& Min & Mean & Max \\
	\hline
	Age (months) & 103 & 180$\pm$39 & 271 \\
	Age (years) & 8.6 & 15$\pm$3.3 & 22.6 \\
	WRAT score & 70 & 102$\pm$15.7 & 145 \\
	\hline
	\end{tabular}
	\label{tab:demographics}
\end{table}

\begin{figure}
    \centering
	\includegraphics[width=8cm]{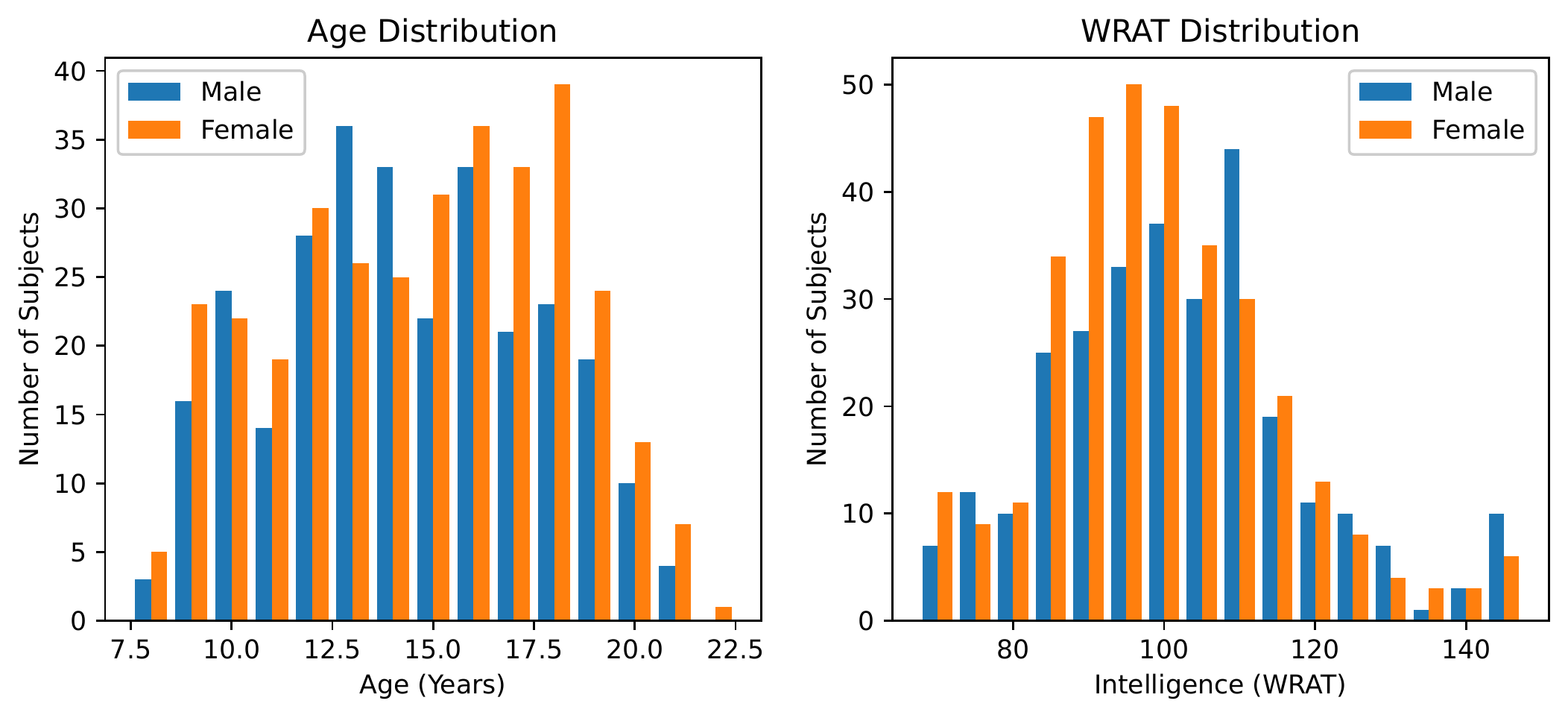}
	\includegraphics[width=5cm]{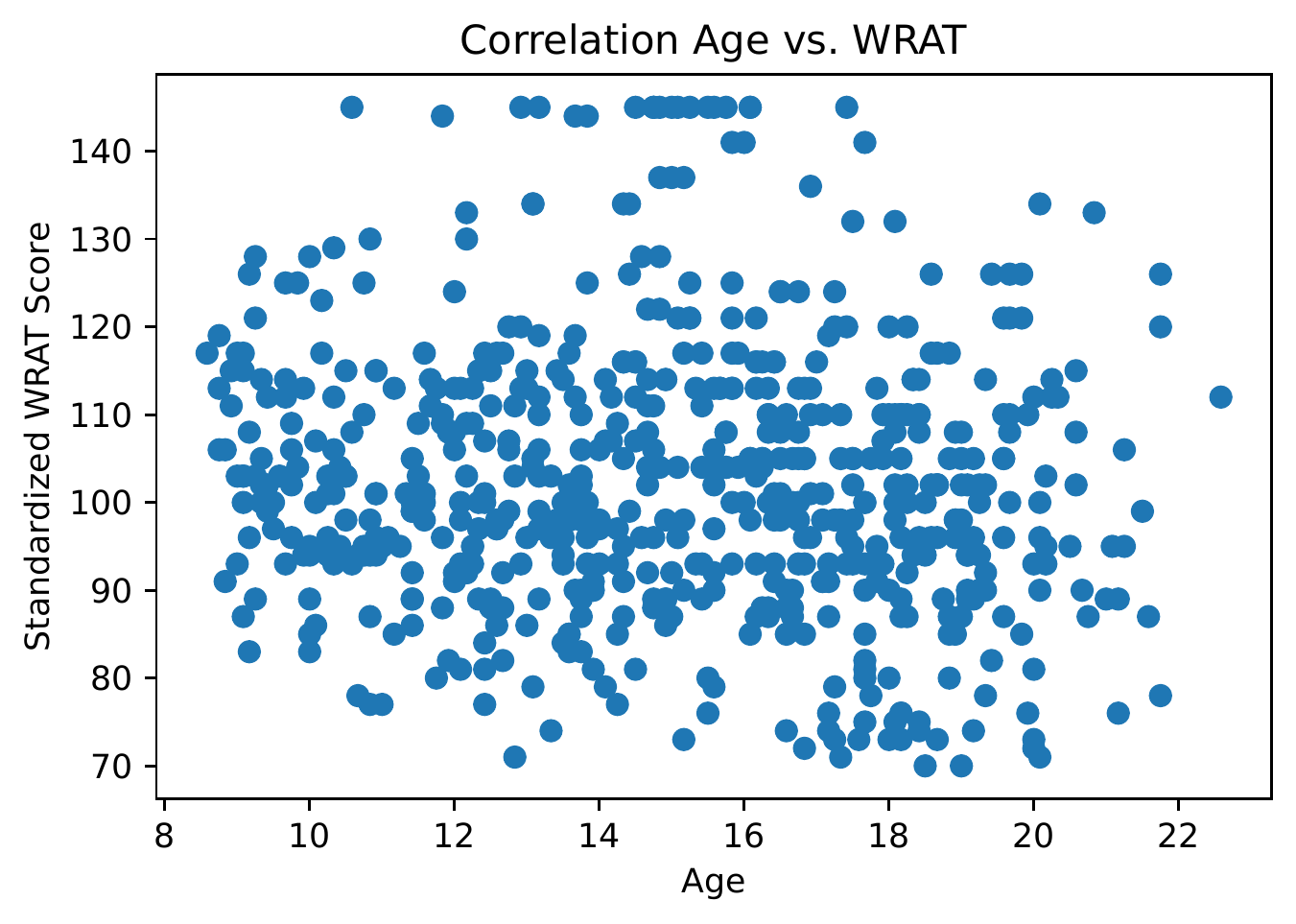}
	\caption{Demographics of the 620-subject subset of the PNC study used in our experiments. \Changes{WRAT score has been adjusted from its raw value by regressing out the effects of age.}} 
	\label{fig:demographics}
\end{figure}

We trained and validated our model on the publicly available PNC dataset. The PNC dataset contains multi-paradigm fMRI data, neurocognitive assessments, and psychiatric evaluations for 1,445 healthy adolescents ages 8-23. We chose 620 subjects from the cohort who had both working memory paradigm (nback) and emotion identification paradigm (emoid) fMRI scans, along with results from the 1-hour Wide Range Achievement Test (WRAT) \cite{Sayegh2014QualityOE} to measure general intelligence.

fMRI was performed using a 3T Siemens TIM Trio whole-body scanner with a single-shot, interleaved multi-slice, gradient-echo, echo-planar imaging sequence. The resolution was set to be 3x3x3 mm with 46 slices. The imaging parameters were TR = 3000 ms, TE = 32 ms, and flip angle = 90 degrees. Gradient magnitude was 45 mT/m, having a maximum slew rate of 200 T/m/s. The duration of the nback scan was 11.6 minutes (231 TR), during which time subjects were asked to conduct the n-back memory task, which is related to working memory and lexical processing \cite{Ragland2002WorkingMF}. The duration of the emoid scan was 10.5 minutes (210 TR), during which time subjects viewed faces displaying different emotions and gave an indication of what emotion was displayed. The demographics of our study cohort are given in Table~\ref{tab:demographics} and the distribution is visualized in Figure~\ref{fig:demographics}.

Data was pre-processed with SPM12\footnote{\url{http://www.fil.ion.ucl.ac.UK/spm/software/spm12/}}. This included using multiple regression for motion correction, as well as spatial normalization and smoothing by a 3mm Gaussian kernel \cite{Friston1995CharacterizingDB}. Pre-processing was similar to \cite{Fang2018-ms}. The Power template \cite{Power2011FunctionalNO} was used to parcellate BOLD signal among 264 regions of interest, from which a $264\times264$ symmetric connectivity matrix was constructed using Pearson correlation. The unique $d=34,716$ entries in the upper right triangle, excluding the main diagonal, were vectorized and taken as the FC features for each subject.

The goal of the experiment was to predict subject age, sex, and intelligence as measured by WRAT score. Prediction performance was measured by root mean squared error (RMSE) for age and intelligence prediction, and accuracy for sex prediction, respectively. \Changes{LatSim was compared against simple linear models (Least Squares and Logistic Regression), a Graph Convolutional Network (GCN), a Multi-Layer Perceptron (MLP), and a Multimodal Graph Convolutional Network (M-GCN).} M-GCN is a recent deep learning model for functional connectome analysis \cite{dsouza2021mgcn} based on the CNN \cite{lecun-gradientbased-learning-applied-1998} architecture. 

\Changes{The inputs to all models were nback, emoid, and the arithmetic sum of nback and emoid task based vectorized FC matrices, from which separate predictions were made and averaged as part of an ensemble. The sum of nback and emoid FC was used to increase ensemble size.  Standardization (Z-score normalization) was performed on the vectorized FC matrices using statistics from the training dataset applied to training, validation, and test datasets. Z-score normalization was performed only for the LatSim model, since the other models sometimes did not converge for Z-score normalized data.} All predictive and feature selection experiments were carried out using 10-fold cross validation (CV), with an 80\% training, 10\% validation, and 10\% test split. \Changes{Hyperparameters were selected using random grid search (see Table~\ref{tab:hyperparameters} for LatSim hyperparameters and Table~\ref{tab:otherhyperparameters} for those of other models). The search grid was initialized to be a 5-decade window around prior assumptions of optimal hyperparameters, with search points occurring at decade intervals for all models. A total of 100 grid points were evaluated with three repetitions. The only exceptions were dropout, which was sampled at 0.1 intervals, latent/hidden dimension, which was set heuristically, and number of training epochs, which was set to just past the maximum best validation epoch for each model individually. Hyperparameters were estimated for the largest training set size ($N=496$) and subsequently used for all training set sizes, with the belief that over-optimization would give a distorted view of models and reduce reproducibility.}

\begin{table}
	\caption{Hyperparameters for PNC experiments (LatSim).}
	\centering
	\begin{tabular}{|l|c|}
		\hline
		Predictive Tasks & Age, Sex, Intelligence\\
		fMRI Paradigms & nback, emoid, nback+emoid \\
		\hline
		Classification Multiplier & $\gamma=1000$ \\
		Sparsity Parameter & $\lambda=10$ \\
		Disentanglement Parameter & $\alpha=100$ \\
		Feature Alignment Parameter & $\beta=0.1$ \\
		\hline
		\Changes{Original Dimension} & \Changes{$d=34,716$} \\
		Latent Dimension & $d'=2$ \\
		Temperature & $\tau=1$ \\
		Feature Dropout Rate & 0.5 \\
		Edge Matrix Dropout Rate & 0.1 \\
		\hline
		Number of Training Epochs & 200 \\
		Optimizer & Adam \\
		Learning Rate & 1e-4 \\
		L2 Regularization Parameter & 1e-4 \\
		Weight Initialization & 1e-4$\cdot\mathcal{N}(0,1)$ \\
		\hline
	\end{tabular}
	\label{tab:hyperparameters}
\end{table}

\begin{table}
    \color{black}
	\caption{Hyperparameters for PNC experiments (comparison models).}
	\centering
	\begin{tabular}{|l|c|}
	    \hline
		Predictive Tasks & Age, Sex, Intelligence\\
		fMRI Paradigms & nback, emoid, nback+emoid \\
	    \hline
	     Model & \emph{Least Squares Regression} \\
	     Implementation & PyTorch\tablefootnote{\url{https://pytorch.org/}}\\
	     \hline
	     Model & \emph{Logistic Regression} \\
	     Implementation & scikit-learn\tablefootnote{\url{https://scikit-learn.org/stable/}} \\
	     Regularization Parameter & C=1 \\
	     \hline 
	     Model & \emph{All Deep Models} \\
	     Optimizer & Adam \\
	     Weight Initialization & PyTorch Default \\
	     Learning Rate & 1e-4 \\
	     \hline
	     Model & \emph{MLP} \\
	     Layers & 34,716 x 100 (hidden) \\
	     Number of Training Epochs & 10,000 \\
	     L1 Regularization Parameter & 1e-2 \\
	     L2 Regularization Parameter & 1e-3 \\
	     \hline
	     Model & \emph{M-GCN} \\
	     Implementation & Github Repository\tablefootnote{\url{https://github.com/Niharika-SD/M-GCN}} \\
	     Number of Training Epochs & 5,000 \\
	     L2 Regularization Parameter & 1e-4 \\
	     \hline 
	     Model & \emph{GCN} \\
	     Layers & 34,716 x 100 (hidden) \\
	     Number of Training Epochs & 10,000 \\
	     Graph Type & Complete \\
	     Neighbor Weight (Total) & 0.5 \\
	     Node Self-loop Weight & 0.5 \\
	     L2 Regularization Parameter & 1e-4 \\
	     \hline
	\end{tabular}
	\label{tab:otherhyperparameters}
\end{table}

\subsubsection{Prediction}

\begin{figure}
    \centering
    \includegraphics[width=8cm]{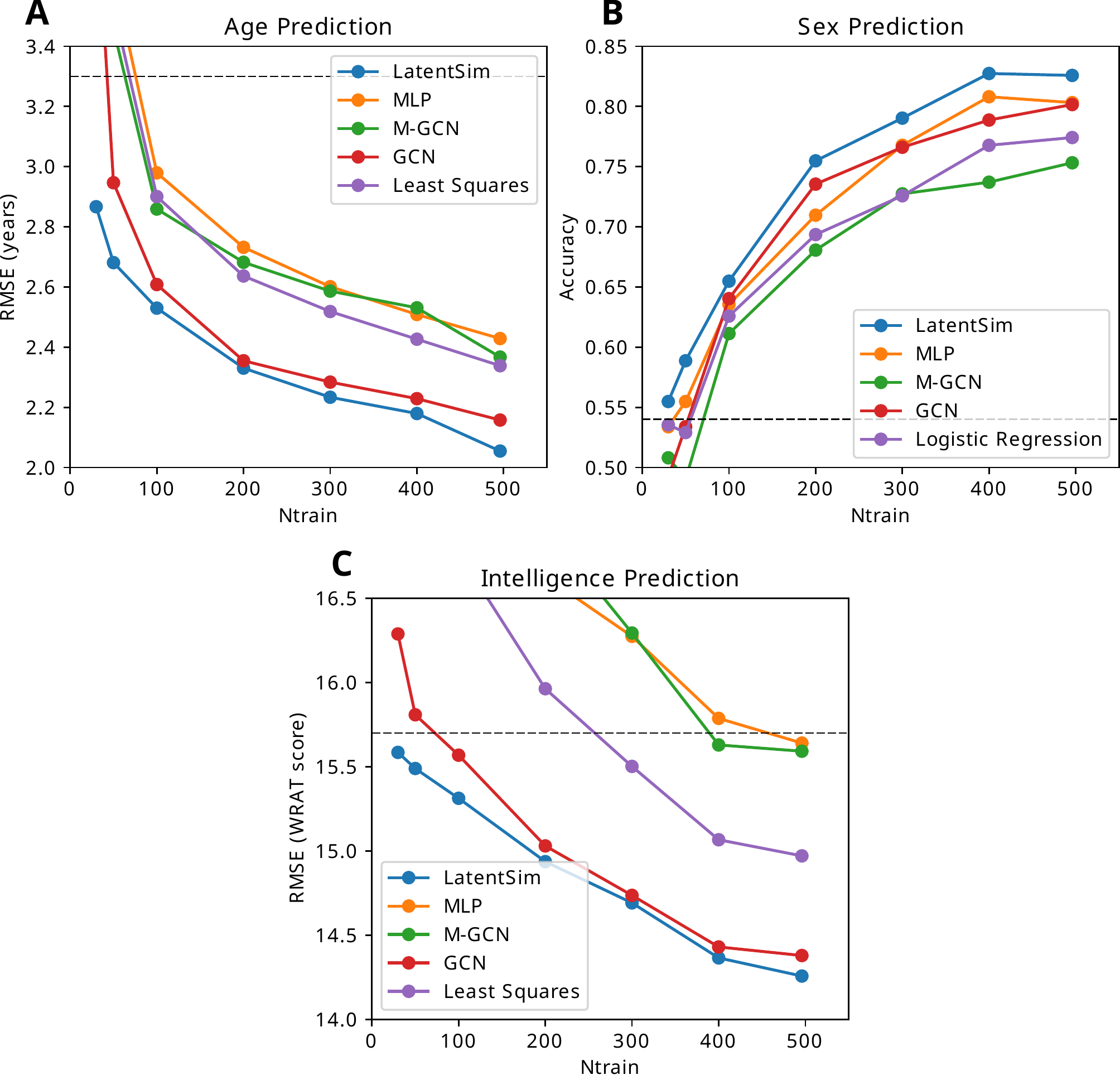}
    \caption{Results of age (\textbf{A}), sex (\textbf{B}), and intelligence (\textbf{C}) prediction experiments on our subset of the PNC dataset. Dashed black lines represent the null model. \Changes{All models except LatSim performed worse than chance at the N=30 training set size for all tasks.}}
    \label{fig:mainresult}
\end{figure}

LatSim achieved superior predictive performance on the PNC dataset in all three predictive tasks, especially at low sample sizes. The result of the entire experiment is given in Figure~\ref{fig:mainresult}, and the low and high sample size results are given in Table~\ref{tab:mainresult}.

At N=30, close to the previously reported threshold of N=36 for modestly reproducible fMRI results, we see that LatSim is the only model not to overfit. It surpassed the other models by a significant margin in two of three predictive tasks. \Changes{Interestingly, LatSim performed much better at small sample sizes than the simple linear models, which we attribute to use of $\mathcal{O}(n^2)$ inter-subject connections rather than the $n$ subjects themselves.  LatSim remains the best performing model until about N=100, at which point it is only slightly better than the other best predictive model, GCN. We note that the GCN model performs almost as well as LatSim, except at low sample sizes. We also note that with a categorical response variable such as sex, the performance of both LatSim and GCN is reduced. We believe the advantage of both LatSim and the GCN model lies in utilizing inter-subject similarities and differences. This is hindered by a lack of granularity in the response variable.}

\begin{table}
	\caption{Results of PNC experiments.}
	\centering
	\begin{tabular}{|l|cc|cc|cc|}
	    \hline
		& \multicolumn{2}{c|}{Age} & \multicolumn{2}{c|}{Sex} & \multicolumn{2}{c|}{Intelligence} \\
		& \multicolumn{2}{c|}{\scriptsize (RMSE, years)} & \multicolumn{2}{c|}{\scriptsize (Accuracy)} & \multicolumn{2}{c|}{\scriptsize (RMSE, WRAT score)} \\
		\hline
		Model & N=30 & N=496 & N=30 & N=496 & N=30 & N=496 \\
		\hline
		Null & 3.3 & & 0.54 & & 15.7 & \\
		M-GCN & 4.47 & 2.37 & 0.51 & 0.75 & 23.27 & 15.59 \\
		MLP & 4.52 & 2.43 & 0.53 & 0.8 & 21.17 & 15.64 \\
		GCN & 3.89 & 2.16 & 0.49 & 0.8 & 16.29 & 14.38 \\
		Linear & 4.36 & 2.34 & 0.54 & 0.77 & 19.8 & 14.97 \\
		LatSim & \textbf{2.86} & \textbf{2.05} & \textbf{0.55} & \textbf{0.82} & \textbf{15.59} & \textbf{14.26} \\
		\hline
		p-value & \textbf{2.2e-6} & \textbf{5e-3} & 0.32 & 0.11 & \textbf{0.02} & 0.30 \\
		\hline
	\end{tabular}
	\label{tab:mainresult}
\end{table}

Based on the prediction results, LatSim can fit a dataset in orders of magnitude less time compared to other models (see Table~\ref{tab:speed}). This makes it possible to perform large-scale bootstrapping, mixture of experts, and ensembling that is not possible with traditional ML models. It also allows for the use of greedy selection.

\begin{table}
	\caption{Training time for all 10 folds of 10-fold cross validation.}
	\centering
	\begin{tabular}{|l|cccccc|}
	    \hline
		Model & LatSim & Lstsq & Logistic & GCN & MLP & M-GCN \\
		\hline
		Epochs & 200 & - & 100 & 1e4 & 1e4 & 5e3 \\
		Training Time & \textbf{4.3s} & \textbf{\textless1s} & 63.4s & 406s & 364s & 5912s \\
		\hline
	\end{tabular}
	\label{tab:speed}
\end{table}

\subsubsection{Significant FCs in prediction}
\label{subsubsec:sigfcs}

\begin{figure}
    \centering
    \includegraphics[width=7cm]{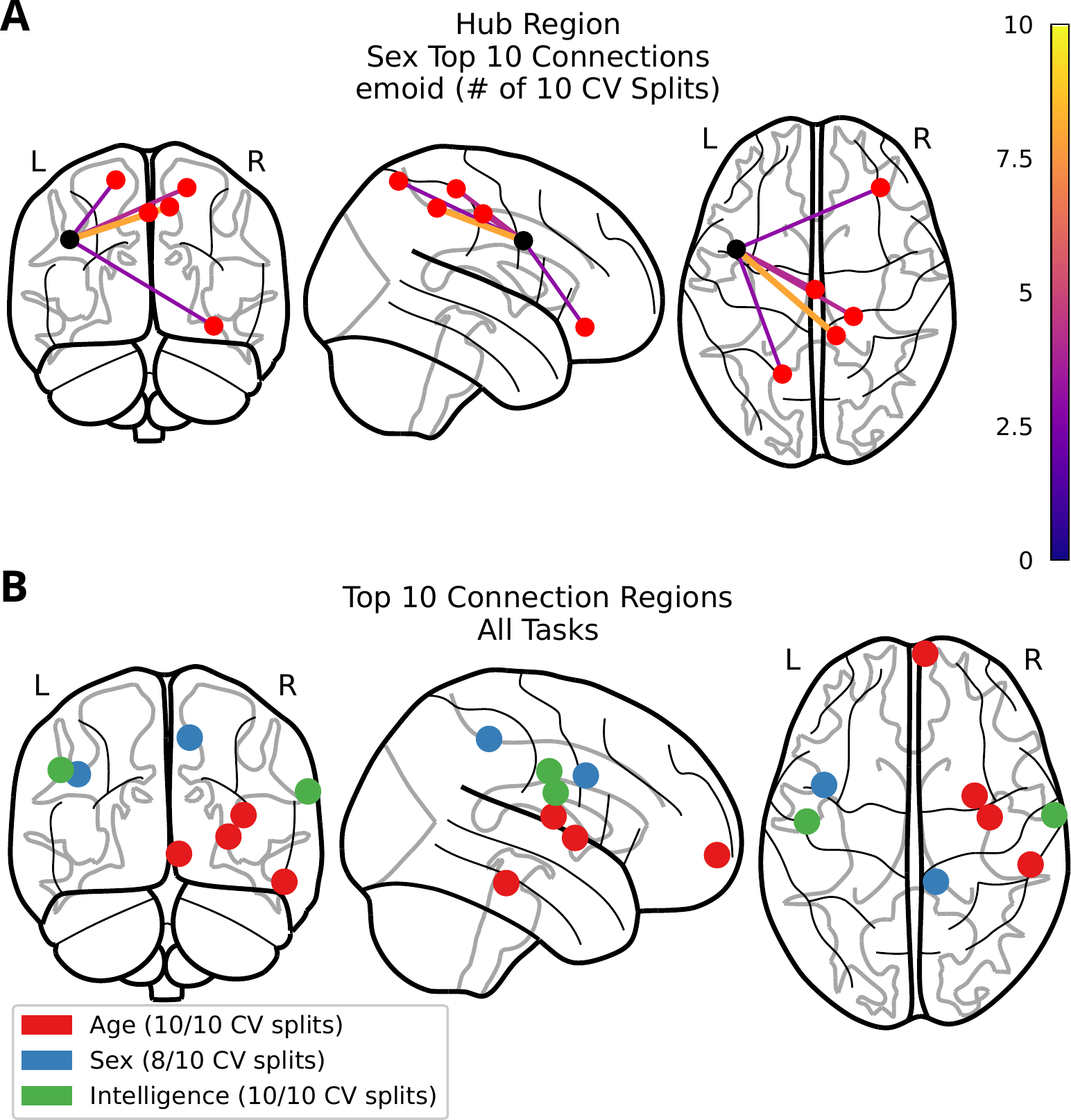}
    \caption{\textbf{A.} Identification of an interesting "hub" region found by emoid paradigm sex prediction that was included in 5 separate connections from among the top 10 connections across all CV splits. \textbf{B.} Visualization of regions found in the top 10 connections of more than 8 CV splits using the greedy selection algorithm.}
    \label{fig:regions}
\end{figure}

\begin{figure}
    \centering
    \includegraphics[width=8cm]{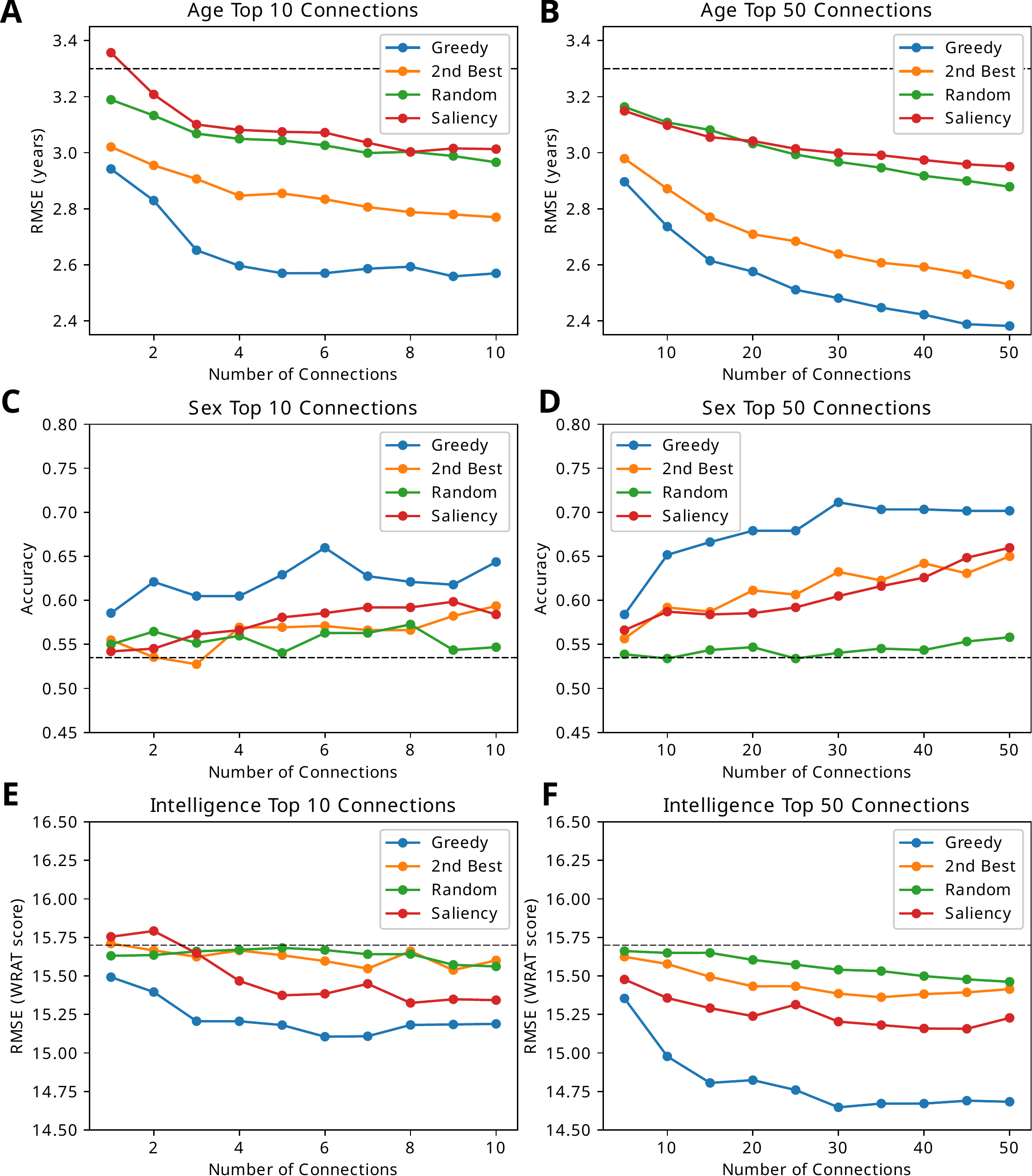}
    \caption{Comparison of four connection selection strategies. Dashed black lines represent the null model. Selection up to 10 connections (\textbf{A}, \textbf{C}, \textbf{E}) was done without dropout, whereas selection up to 50 connections (\textbf{B}, \textbf{D}, \textbf{F}) was done with 0.5 dropout.}
    \label{fig:selection}
\end{figure}

\begin{table*}
	\caption{Most important connections for discriminating age, sex, and intelligence among healthy adolescents. The \# CV splits column shows the number of CV splits for which the connection appeared in the top 10 connections of the greedy selection algorithm.}
	\centering
	\footnotesize
	\begin{tabular}{|ccc|ccc|c|c|c|}
	    \hline
		Region 1 & MNI Coords & Network & Region 2 & MNI Coords & Network & \# CV Splits & Paradigm & Prediction Task \\
		\hline
		Insula\_R & (36,-9,14) & SMT & Putamen\_R & (29,1,4) & SUB & 10/10 & Both & Age \\
		Temporal\_Inf\_R & (55,-31,-17) & UNK & Frontal\_Med\_Orb\_R & (6,67,-4) & DMN & 10/10 & Both & Age \\
		Frontal\_Mid\_L & (-34,55,4) & FRNT & Frontal\_Mid\_Orb\_L & (-42,45,-2) & FRNT & 9/10 & nback & Age \\
		Thalamus\_R & (6,-24,0) & SUB & Left Brainstem & (-5,-28,-4) & SUB & 9/10 & emoid & Age \\
		\hline
		Precentral\_L & (-41,6,33) & FRNT & Temporal\_Pole\_Mid\_R & (11,-39,50) & SAL & 8/10 & emoid & Sex \\
		Insula\_R & (27,16,17) & UNK & Frontal\_Inf\_Orb\_R & (49,35,-12) & DMN & 3/10 & emoid & Sex \\ 
		Temporal\_Pole\_Mid\_R & (46,16,-30) & DMN & Temporal\_Pole\_Mid\_R & (52,7,-30) & DMN & 3/10 & nback & Sex \\
		Frontal\_Sup\_Orb\_R & (24,32,-18) & UNK & Fusiform\_R & (27,-37,-13) & DMN & 3/10 & nback & Sex \\
		\hline
		Postcentral\_L & (-49,-11,35) & SMT & Postcentral\_R & (66,-8,25) & SMT & 10/10 & Both & Intelligence \\
		Temporal\_Mid\_R & (52,-2,-16) & DMN & Precuneus\_R & (10,-62,61) & DRSL & 6/10 & nback & Intelligence \\
		Cerebelum\_6\_L & (-16,-65,-20) & CB & Postcentral\_R & (66,-8,25) & SMT & 5/10 & emoid & Intelligence \\
		Precentral\_R & (44,-8,57) & SMT & Temporal\_Inf\_L & (-42,-60,-9) & DRSL & 5/10 & emoid & Intelligence \\
		\hline
	\end{tabular}\\
	\vspace{0.2cm}
	\raggedright
	SMT=Sensory/Somatomotor, CNG=Cingulo-opercular Task Control, AUD=Auditory, DMN=Default Mode, MEM=Memory Retrieval, VIS=Visual, FRNT=Fronto-parietal Task Control, SAL=Salience, SUB=Subcortical, VTRL=Ventral Attention, DRSL=Dorsal Attention, CB=Cerebellum, UNK=Uncertain
	\label{tab:connections}
\end{table*}

The most important FCs for all prediction tasks are given in Table~\ref{tab:connections}. All connections are given with Automated Anatomical Labeling (AAL) region names \cite{ROLLS2020116189} and with Montreal Neurological Institute (MNI) region coordinates. For age prediction, the most important connections were Insula\_R to Putamen\_R and Temporal\_Inf\_R to Frontal\_Med\_Orb\_R, being present in the top 10 connections for both the nback and emoid paradigms. For sex prediction, the Precentral\_L to Temporal\_Pole\_Mid\_R FC was found in the top 10 connections for the emoid paradigm. For intelligence prediction, the Postcentral\_L to Postcentral\_R FC was found in the top 10 connections for both the nback and emoid paradigms.  In addition, for sex prediction, we identified the Left Inferior Frontal Gyrus (Precentral\_L) as a region making multiple top 10 connections, as shown in Figure~\ref{fig:regions}.

Using only the first few connections gives half of the predictive power of using the full set of $d=34,716$ connections. In particular, Figure~\ref{fig:selection} shows that the first 3 connections, if properly chosen, can contain more information than the next 50 connections, chosen in the same manner. Specifically, 10 FCs can explain 21\% of variance for age, 50 FCs can explain 27\%, whereas with the full set of FCs the GCN model can explain 35\% and LatSim can explain 38\%. The selected connections were chosen using the greedy feature selection algorithm. Figure~\ref{fig:selection} shows that the FCs chosen by greedy selection are superior to those chosen by gradient-based saliency, as well as to random FCs. Additionally, we compared the FCs chosen by greedy selection to the next-best FCs that would be chosen by it. We believe this helps validate the significance of our identified connections, since, for small numbers of connections, we could not find a minimal combination of FCs that performed as well as that found by greedy selection.

\begin{figure*}
    \centering
    \includegraphics[width=18cm]{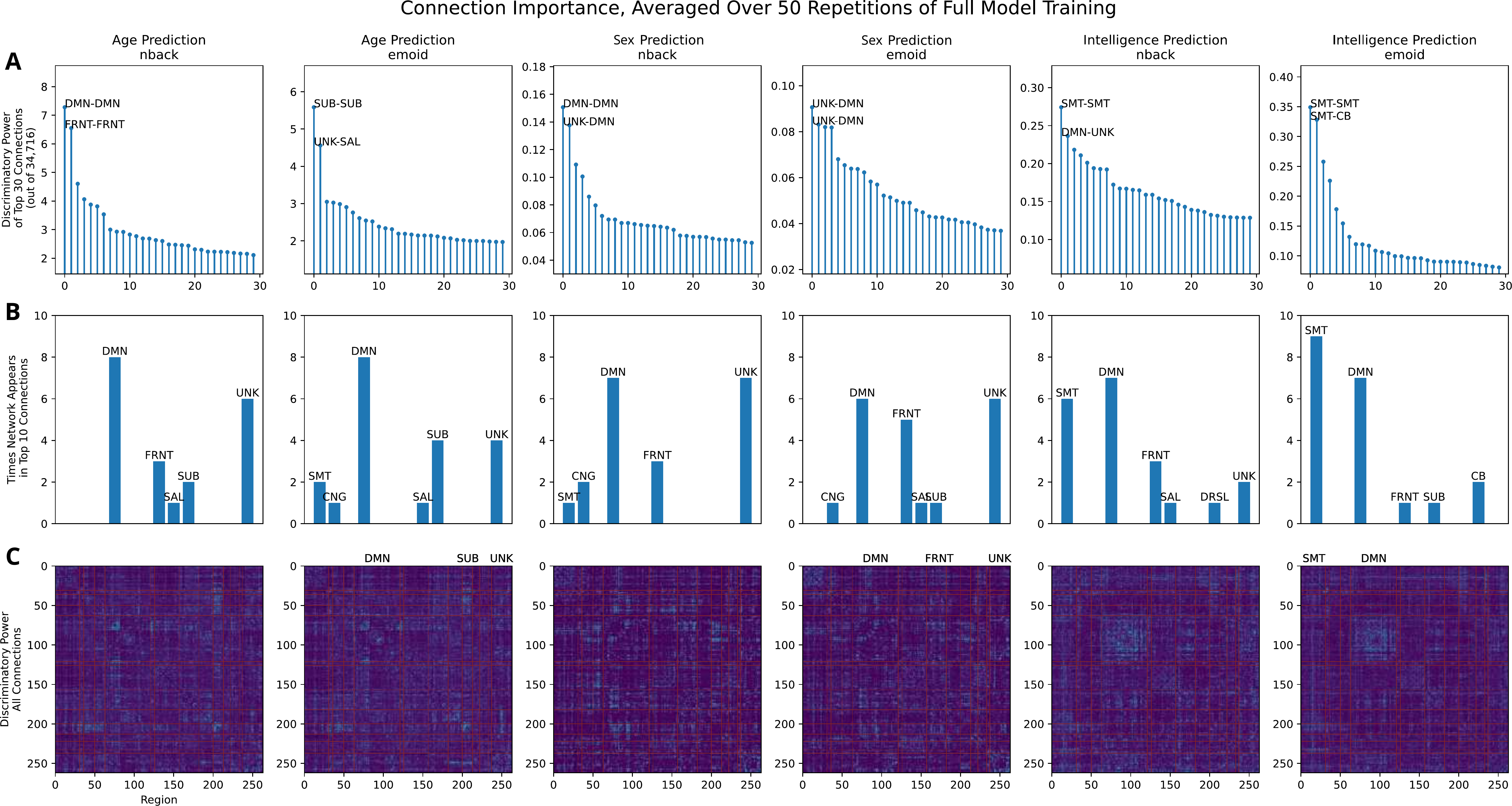}
    \caption{Important connections identified by running the full model with the entire $d=34,716$ set of connections as inputs. As with greedy selection, we show that the first several connections are far more important than the remaining ones (\textbf{A}). Notably, the DMN is highly represented in the top 10 connections for all predictive tasks and modalities (\textbf{B}). The DMN as a whole seems to be important for intelligence prediction (\textbf{C}). Importance was averaged over 50 repetitions of an 80-10-10 train/validation/test split. Discriminative power was calculated as in Equation~\ref{eq:select}. Correlation was greater than zero for all connections. See Table~\ref{tab:connections} for definitions of abbreviations.}
    \label{fig:bigimportant}
\end{figure*}

Selecting connections with the fully trained LatSim model corroborated the trend found by greedy selection. As seen in Figure~\ref{fig:bigimportant}, we identified a very few "core" connections that were disproportionately important to the prediction task. The rest of the connections were interchangeable in terms of discriminative ability. Note, for instance, the rapid increase in accuracy for the 3 best FCs and the subsequent plateau in Figure~\ref{fig:selection}. Likewise, almost all of the connections found in the top 50 connections by greedy selection were also found in the top 50 connections of the full model.

\section{Discussion}
\label{sec:discussion}

\subsection{Significant functional networks}
\label{subsec:networks}

The top connections identified by this study contain regions that fall into the default mode (DMN), subcortical (SUB), fronto-parietal task control (FRNT), and sensory/somatomotor (SMT) brain functional networks (FNs). Abbreviations are given as a footnote to Table~\ref{tab:connections}. Regions that belong to the same FN (within-module) tend to be more synchronized than regions from different FNs (between-module) \cite{10.1093/cercor/bhu036}. In Figure~\ref{fig:bigimportant}C, blocks on the main diagonal of the FC matrices represent connections within-module, while blocks off the main diagonal represent connections between-module. Recently, \textit{Jiang et al.} found that, in an older population, connections between the DMN, SMT, and SUB networks were highly predictive for age  \cite{https://doi.org/10.1002/advs.202201621}. They also found that a DMN-SUB connection was correlated with high cognitive performance.

The DMN was overrepresented in the top 10 connections for all predictive tasks; 36\% of regions identified were part of the DMN, whereas DMN regions constitute 22\% of the Power atlas. Robust developmental changes have been identified in the DMN, and DMN connectivity has been positively correlated with high cognitive performance \cite{PERSSON20142107}. \textit{Fan et al.} found that DMN connectivity increases from childhood until young adulthood \cite{FAN2021117581}. \textit{Pan et al.} identified FCs which included DMN regions to be more important in predicting intelligence than FCs which didn't \cite{9311755}. 

The SMT network was overrepresented in top 10 connection regions for intelligence prediction. In that task, 43\% of top 10 connection regions belonged to the SMT network, whereas SMT regions constitute 13\% of the Power atlas. It is known that dysfunction in the SMT network is correlated with depression \cite{ZHANG2021118188}. However, FC represents synchronization between brain regions, and the cause of altered FC may not lie in the region itself. Table~\ref{tab:connections} shows that the top SMT connections involve the CB network, leading to the idea that complex motor control is related to intelligence.

Many of the most important connections we identified for each predictive task are not recognized as part of an FN, and are classified as unknown-network (UNK). 24\% of regions identified in top 10 connections are labeled UNK, whereas UNK regions constitute 10\% of ROIs in the Power atlas. These connections include cerebellar regions; some cerebellar regions are not included in the CB network because they contribute to functions other than motor function \cite{pmid32888303}, including social thinking and emotion \cite{pmid32888303}. \textit{Zhang et al.} recently found disrupted effective connectivity in UNK cerebellar regions in individuals with schizophrenia, relative to controls \cite{pmid35842099}.

\subsection{Significant FCs}
\label{subsec:connections}

Greedy selection identified 4 FCs present in more than 8 out of 10 CV splits for one of the predictive tasks: 

\begin{itemize}
\item \textbf{Insula\_R to Putamen\_R (Age)}. The Insula\_R has many functions in humans dealing with low-level sensation, emotion, and high-level cognition \cite{pmid28644199}. \textit{Mazzola et al.} hypothesized that the Insula\_R participates in the social brain and found increased activation when participants watched scenes of joyful or angry actors \cite{pmid27375449}. Increased Putamen\_R volume has been linked to autism spectrum disorder \cite{10.3389/fnhum.2014.00957}, and reduced amygdala-Putamen\_R FC has been linked to ADHD \cite{MCLEOD2014566}.
\item \textbf{Temporal\_Inf\_R to Frontal\_Med\_Orb\_R (Age)}. The Temporal\_Inf\_R region is associated with language processing \cite{pmid15809000}. Temporal\_Inf\_R FC was found to be decreased in adolescent schizophrenia patients \cite{pmid34997425}. The Frontal\_Med\_Orb\_R region is part of the prefrontal cortex and is associated with dysfunctional connectivity in major depressive disorder \cite{pmid30813750}.
\item \textbf{Precentral\_L to Temporal\_Pole\_Mid\_R (Gender)}. The Precentral\_L region is associated with reading and language processing \cite{pmid21237182}. \textit{Delvecchio et al.} found morphological differences in this region between sexes \cite{pmid33613268}. The Temporal\_Pole\_Mid\_R region is linked to social contracts, precautions, and strategies \cite{pmid23548839}.
\item \textbf{Postcentral\_L to Postcentral\_R (Intelligence)}. This is a connection between regions symmetric about the body mid-line. The postcentral gyrus is involved in proprioception and contains the primary somatosensory cortex. Lesions in these regions may cause speech dysfunction \cite{pmid22130092}\cite{HardCitation}. \textit{Sander et al.} found inter-hemispheric connectivity to play a role in the ability to learn new languages \cite{10.1093/cercor/bhac131} .
\end{itemize}

Notably, AAL regions extend over a large area, and Power atlas ROIs do not correspond exactly to AAL regions.

\subsection{Robustness to spurious correlation}
\label{subsec:robustness}

\begin{figure}
    \centering
	\includegraphics[width=8cm]{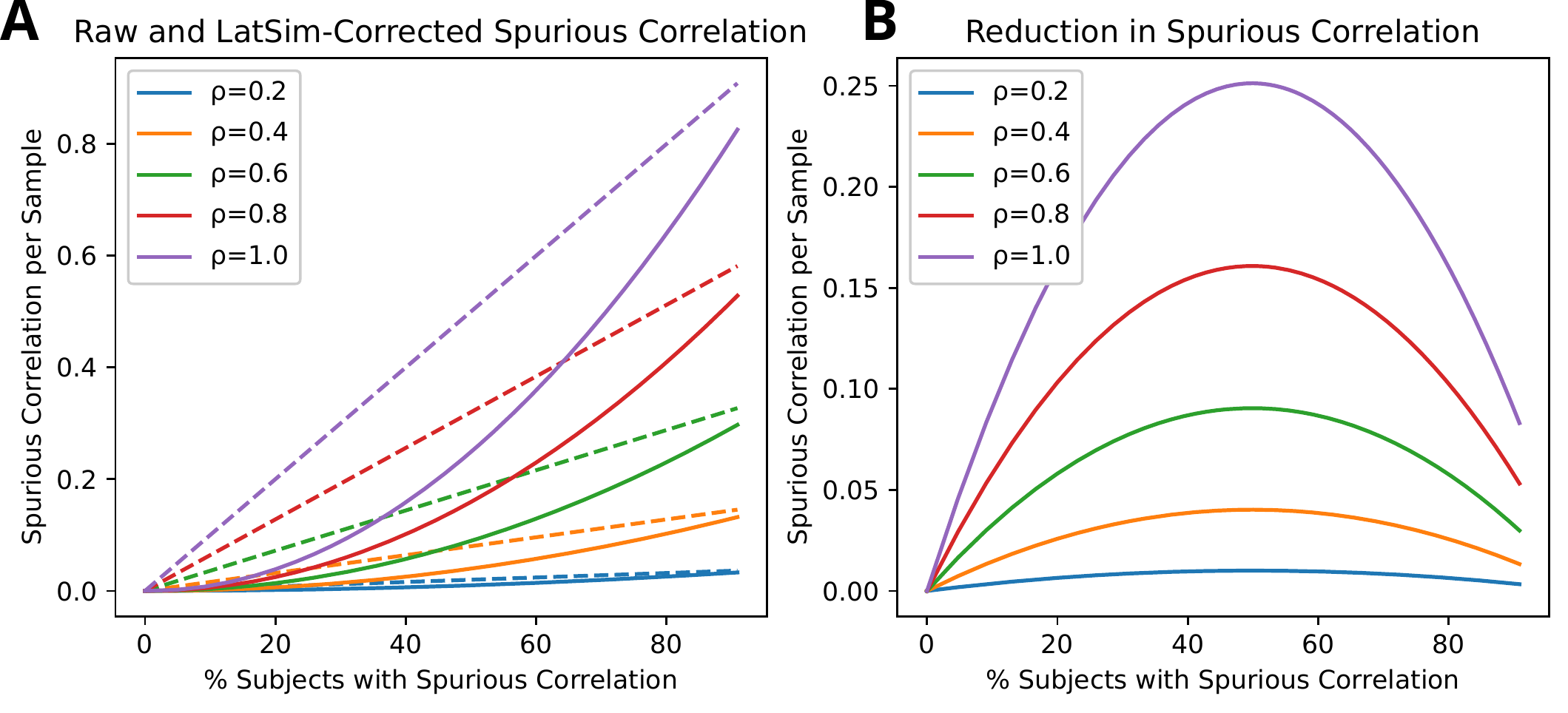}
	\caption{\textbf{A.} Spurious correlation per sample in a traditional ML model (dashed lines) versus LatSim (solid lines). \textbf{B.} The absolute reduction in spurious correlation as a function of frequency in the sample.} 
	\label{fig:theory}
\end{figure}

\Changes{
In this section, we argue that LatSim is robust to spurious correlation because it identifies features based on $\mathcal{O}(n^2)$ inter-subject connections, rather than the number of subjects in the cohort.}

\Changes{
Assume feature $X$ is spuriously correlated with response variable $Y$ on a subset of the cohort $S\subseteq C$, $s=|S|$, $n=|C|$, and $X,Y\sim\mathcal{N}(0,1)$. That is, for each subject $u$:}

\begin{equation}
    |\rho_S| \begin{cases}
    > 0 & u \in S \\
    \approx 0 & u \in C \setminus S
    \end{cases}
\end{equation}

\Changes{
LatSim uses weighted inner product similarity $wX_1wX_2$ between the features of two subjects as input, where $w$ is a learned weight. The correlation between $wX_1wX_2$ and $D=(Y_1-Y_2)^2$ determines how well this feature pair predicts the response variable: }

\begin{equation}
    \begin{split}
        \rho_{XX,D} &= \frac{\sigma^2_{XX,D}}{\sqrt{\sigma^2_{XX}\sigma^2_D}} \\
        &= \frac{\text{Cov}[wX_1wX_2,(Y_1-Y_2)^2]}{\sqrt{\text{Var}[wX_1wX_2]\text{Var}[(Y_1-Y_2)^2]}} \\
        &= \frac{\mathbb{E}[X_1X_2(Y_1-Y_2)^2-\mu_{XX}\mu_{D}]}{\sqrt{\mathbb{E}[X^2_1X^2_2-\mu^2_{XX}]\mathbb{E}[(Y_1-Y_2)^4-\mu^2_D]}} \\
        &= \frac{\mathbb{E}[X_1X_2(Y_1-Y_2)^2-0\cdot 2]}{\sqrt{\mathbb{E}[X^2_1X^2_2-0^2]\mathbb{E}[(Y_1-Y_2)^4-2^2]}} \\
        &= \frac{\mathbb{E}[X_1X_2Y^2_1-2X_1X_2Y_1Y_2+X_1X_2Y^2_2]}{\sqrt{(1\cdot1)(12-4)}} \\
        &= \frac{0-2\rho^2_{XY}+0}{\sqrt{8}} \\
        &= \begin{cases}
        -\frac{\rho^2_S}{\sqrt{2}} & 1,2 \in S \\
        0 & \text{otherwise}
        \end{cases}
    \end{split}
\end{equation}

\Changes{
Since expectation is a linear operator, we can find the average value over the entire cohort:
}

\begin{equation}
    \label{eq:latsimcorr}
    \begin{split}
        \rho_{XX,D} &= -\frac{s(s-1)}{n(n-1)}\frac{\rho^2_S}{\sqrt{2}} \approx -\frac{s^2}{n^2}\frac{\rho^2_S}{\sqrt{2}} \\
        &\approx k\frac{s^2}{n^2}\rho^2_S
    \end{split}
\end{equation}

\Changes{
Conversely, in a traditional model, feature $X$ is correlated with response variable $Y$ as the size of the subset $S$:
}

\begin{equation}
    \label{eq:traditionalcorr}
    \rho_{X,Y} = \frac{s}{n}\rho_S
\end{equation}

\Changes{
A plot of the functions in Equations~\ref{eq:latsimcorr} and \ref{eq:traditionalcorr} is given in Figure~\ref{fig:theory}. The maximum reduction in spurious correlation occurs at $s/n=0.5$ and is about 1/4 of the value of the spurious correlation. The relative reduction is linear and maximal when $s=0$, i.e., there are no subjects with spurious correlation (not shown). As $s/n$ increases, the reduction in spurious correlation is diminished. This suggests that large model capacity is not the only reason complicated models falter at low sample sizes. We see in our experiments, e.g., in Table~\ref{tab:mainresult}, that the linear models perform worse than both LatSim and some other deep learning models.}

\Changes{
Like LatSim, a k-layer GNN model also works on interactions between subjects, but as an adjunct to the prediction from the node self-loop. It also requires either additional degrees of freedom to estimate edge weights, or an arbitrary choice of a distance function and/or threshold.} We believe the reason that a GCN model did so well in our experiments is that we made it incredibly simple: only 2 layers were used, and edge weights were uniform and equal in sum to the self-weights. It was found that expanding the GCN to 3 or 4 layers hurt performance. We believe the performance benefit comes from having a good prior and feature selection, not additional model capacity. Due to the very weak relationships between features and response variables in our data, we believe the advantage of the GCN was in averaging. This strategy breaks down at low sample sizes, where spurious feature correlation still causes large errors to be present at the node self-loop.

\section{Conclusion}
\label{sec:conclusion}

This paper proposes a novel model, LatSim, in the vein of metric learning, that is robust against overfitting at small sample sizes. It is interpretable, computationally efficient, multi-task and multi-view capable, and able to enforce feature disentanglement. First, we showed that LatSim is superior in the small sample size, high dimensionality regime, through both simulation and experiments on real datasets. Second, we identified specific connections within and between the sensory/somatomotor, default mode, fronto-parietal task control, and subcortical networks that are highly discriminative for age, sex, and intelligence in healthy adolescents. Third, we quantified the number of features required to attain a given prediction accuracy. Fourth, we showed that there are several core connections that are more discriminative for each predictive task than other connections. Finally, we found that connections identified by greedy selection were superior compared to those found by saliency methods. Our model may spur new research into algorithm development and, in turn, lead to new insights into the mechanisms underlying human cognition.

\bibliographystyle{IEEEtran}
\bibliography{latsim.bbl}

\end{document}